# General Effect Modelling (GEM)

## Part 3. GEM applied on proteome data of cerebrospinal fluid of multiple sclerosis and clinically isolated syndrome

Mosleth, E.F.[1], Myhr, K.-M. [2,3], Vedeler,C.A.[2,3], Berven,F.S. [4], A.Lysenko[5], Gavasso,S. [2,3], Liland, K.H.[6]

[1] Nofma AS, Norwegian Institute of Food, Fisheries and Aquaculture Research, Osloveien 1, 1430 Ås, Norway.
[2] Department of Clinical Medicine, University of Bergen, Norway.
[3] Neuro-SysMed, Department of Neurology, Haukeland University, Hospital, Bergen, Norway. 5
[4] Proteomics Unit (PROBE), Department of Biomedicine, University of Bergen, Bergen, Norway.
[5] Science Mathematics, RIKEN Center for Integrative Medical Sciences, Yokohama, Japan. 9, Center for Cancer
[6] Faculty of Science and Technology, Norwegian University of Life Sciences, 1430 Ås, Norway. 6

## Abstract

The novel data analytical platform General Effect Modelling (GEM), is an umbrella platform covering different data analytical methods that handle data with multiple design variables (or pseudo design variables) and multivariate responses. GEM is here demonstrated in an analysis of proteome data from cerebrospinal fluid (CSF) from two independent previously published datasets, one data set comprised of persons with relapsing-remitting multiple sclerosis, persons with other neurological disorders and persons without neurological disorders, and one data set had persons with clinically isolated syndrome (CIS), which is the first clinical symptom of MS, and controls. The primary aim of the present publication is to use these data to demonstrate how patient stratification can be utilised by GEM for multivariate analysis. We also emphasize how the findings shed light on important aspects of the molecular mechanism of MS that may otherwise be lost. We identified proteins involved in neural development as significantly lower for MS/CIS than for their respective controls. This information was only seen after stratification of the persons into two groups, which were found to have different inflammatory patterns and the utilisation of this by GEM. Our conclusion from the study of these data is that disrupted neural development may be an early event in CIS and MS.

## Introduction

Multiple sclerosis is a serious inflammatory and degenerative disease in the central nervous system (CNS) [1]. The present study investigates a dataset of early signatures of multiple sclerosis, called clinically isolated syndrome (CIS), and a dataset of persons with and without relapsing-remitting multiple sclerosis, the most common form of multiple sclerosis. CIS is the first clinical attack, that may or may not develop to further dissemination in space and time, which is the key criterion of the diagnosis of multiple sclerosis. Relapsing-remitting multiple sclerosis is characterized by the repetition of brief episodes of neurologic dysfunction sustained by acute CNS inflammation and the synchronous appearance of demyelinating lesions. For simplicity, we use the term MS and not relapsing-remitting MS in this publication.

Recent data indicate that MS is caused by an initial infection by Epstein-Barr virus [2, 3] – but the risk of developing MS is modified by a complex interplay of environmental and genetic factors [2-11]. The course of the MS is heterogeneous, and the diagnostic definition of MS has changed over time [8].

In the present publication, we apply the data analysis General Effect Modelling (GEM) on CSF proteomics data. GEM is an umbrella covering different data analytical methods that handle data with multiple design variables (or pseudo-design variables) and multivariate responses. GEM covers the method Effect plus Residual (ER) modelling, which we previously presented and applied to the same data. In the present study, we extend the previous analysis by using GEM which is more flexible and also allows continuous design variables.

The primary aim of this publication is to demonstrate in detail the GEM methodology and show how GEM can utilise patient stratification. We also discuss the detected proteome changes to highlight that GEM contributes to shed light on underlying molecular mechanisms of the disease that may otherwise be lost.

## Material and method

The CSF proteome was analysed from two different datasets: a dataset of CIS and a dataset of MS.

### Dataset 1. The MS dataset

The MS dataset consisted of CSF proteomics data from persons with MS, other neurological disorders and without neurological disease who had undergone spinal puncture for other reasons. All non-MS persons were considered as controls. In the first publication of these data, explorative cluster analysis stratified all persons into two groups [12], later called group A and group B [13] as presented in **Table 1**, where group B was found to have elevated inflammatory CSF pattern compared with group A. This included most IgGs, which were elevated in group B compared with group A. By confidence intervals within nonMS and within MS, as much as 1/3 of the proteins were differentially expressed in group B compared with group A [13].

As the majority of persons without MS do not have inflammatory patterns in CSF, most nonMS were in group A, and vice versa most persons with MS were in group B.

GEM was applied using four design variables considering the experiment as a four-way full factorial design (**Table 1**). The design variables were denoted d1-d4 as follows:

   d1: Two-level variable of group identity ('grA' for group A or 'grB' for group B)

   d2: Two-level variable of disease status ('nonMS' or 'MS')

   d3: Two-level variable of gender ('F' for Females or 'M' for Males)

   d4: Continuous variable: age at Lumbar Puncture

The age of the lumbar puncture spanned from 18 to 51 years.

*Table 1. MS dataset. The number of persons in each combination of group (group A and group B), MS diagnosis (nonMS and MS) and gender (Females and Males).*

|  |  | Group A | Group B | Tot |
|---|---|---|---|---|
| **nonMS** | **Females** | 32 | 7 | 39 |
|  | **Males** | 23 | 2 | 25 |
| **MS** | **Females** | 6 | 22 | 28 |
|  | **Males** | 1 | 8 | 9 |

### Dataset 2, the CIS dataset

The CIS dataset consisted of CSF proteomics data from dataset patients where the proteome was analysed close to the first clinical attack (< 2 months). The CSF proteome of CIS was compared with control persons who did not have any neurological disorders. Some of the CIS had elevated CSF inflammation, while others did not show elevated CSF inflammation signature. Two persons were unknown for CSF inflammatory markers and omitted here. Analogously to group A in the MS dataset, we defined group A as persons who did not display elevated CSF inflammation. Persons with elevated CSF inflammation were defined as group B. None of the non-CIS were in group B in this dataset

The frequency of CIS, controls and gender in group A are presented in **Table 2**. The age of the first clinical episode suggestive of MS spanned from 18 to 50 years for all CIS.

*Table 2. CIS dataset. Number of persons in each combination of CIS diagnosis (non-CIS and CIS) and gender (females and males) for persons in group A.*

|         |   | Group A | Group B | Tot |
|---------|---|---------|---------|-----|
| non-CIS | F | 20      | 0       | 20  |
|         | M | 25      | 0       | 25  |
| CIS     | F | 11      | 22      | 33  |
|         | M | 6       | 6       | 12  |

### GEM

In an experiment with multiple design variables, GEM isolates the effects of each design variable and includes the residual of the complete model for validation. Here we used GEM to analyse separately the effects of one design variable at the time.

A mathematical description of GEM is given in **Figure 1** as illustrated for two design variables (d1 and d2), and r codes are given in the Appendix. GEM consists of two steps where the first step is a general linear model that is only used to estimate effects. The effects of each design variable plus the residuals are called ER values. The second step covers multivariate and univariate analysis of the ER values of each design variable. We here present the results of multivariate analysis by the supervised multivariate analysis Partial Least Square [14] within the GEM framework (GEM-PLS). For more details on GEM, we refer to Part 1 of this publication series.

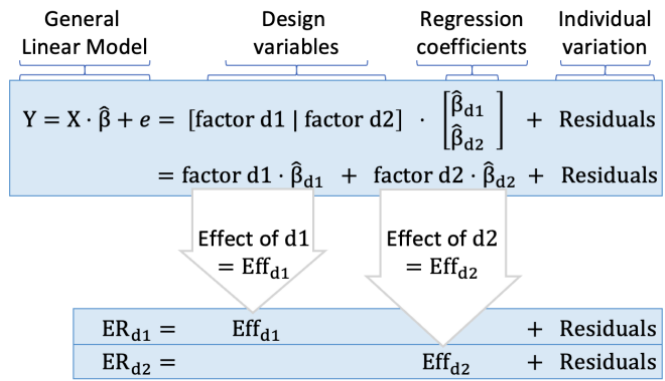

*Figure 1*. Mathematical description of GEM shown for a study with two design variables (d1 and d2). The first step of GEM is the estimation of the effects of the design variables using a General Linear Model (GLM). X is the design matrix where d1 and d2 are columns. Each design variable was non two levels: d1 is 'group A' or 'group B' and d2 is the disease status: 'nonMS' and 'MS'. The regression coefficients ($\hat{\beta}$'s) are estimated from the model, and the residuals contain the variation that can not be explained by the two design factors. The effects of one design variable at a time along with the residuals of the original GLM model are used in step 2 of GEM which can be any multivariate or univariate statistical analysis of the ER values. Mathematically described, the effects of one design variable at a time are included in the statistical analyses, whereas the effects of the other design variables are orthogonalized and omitted in the statistical analyses. The degrees of freedom consumed in the first step must be taken into consideration in the following analysis when applying data analytical methods that consider degrees of freedom.

## Results and discussion

**Raw data plot of a subset of the proteins**

We here first pay attention to the complement proteins which were found in our former publication of these data to play an important role in MS [13].

Univariate analysis by confidence interval of the complement proteins applied within each group of the two datasets showed that most complement proteins had mean expression levels lower for MS/CIS than for nonMS as considered within group (**Figure 2**). This pattern was found both for the MS dataset and for the CIS dataset.

When considering all four combinations of group and disease status (**Figure 3**), the complement proteins C1RL, C2, C3, C5, C6, C7, C8A, C8B, C8G, C9, CFB, CFH, CFI displayed a similar pattern of variation. Within each group, MS had lower mean expression levels of these proteins than nonMS. When comparing the groups, group B (the group with high CSF inflammation) had higher expression levels than group A.

In **Figure 3**, the most frequent combinations are marked with circles. As the majority of persons without MS do not have elevated CSF inflammation, group A is the most frequent among nonMS. In contrast, the majority of MS patients have elevated CSF inflammation with elevated expression of most IgGs, group B MS is therefore most frequent among MS. Thus, the two most frequent combinations are: "group A nonMS" and "group B MS". If "group B MS" is compared with "group A nonMS", there is a confundation between group identity and disease status. The lower levels of the complement proteins within group would therefore be overseen if the persons had not been stratified into two groups. This was also emphasised in the original publication of this material [12].

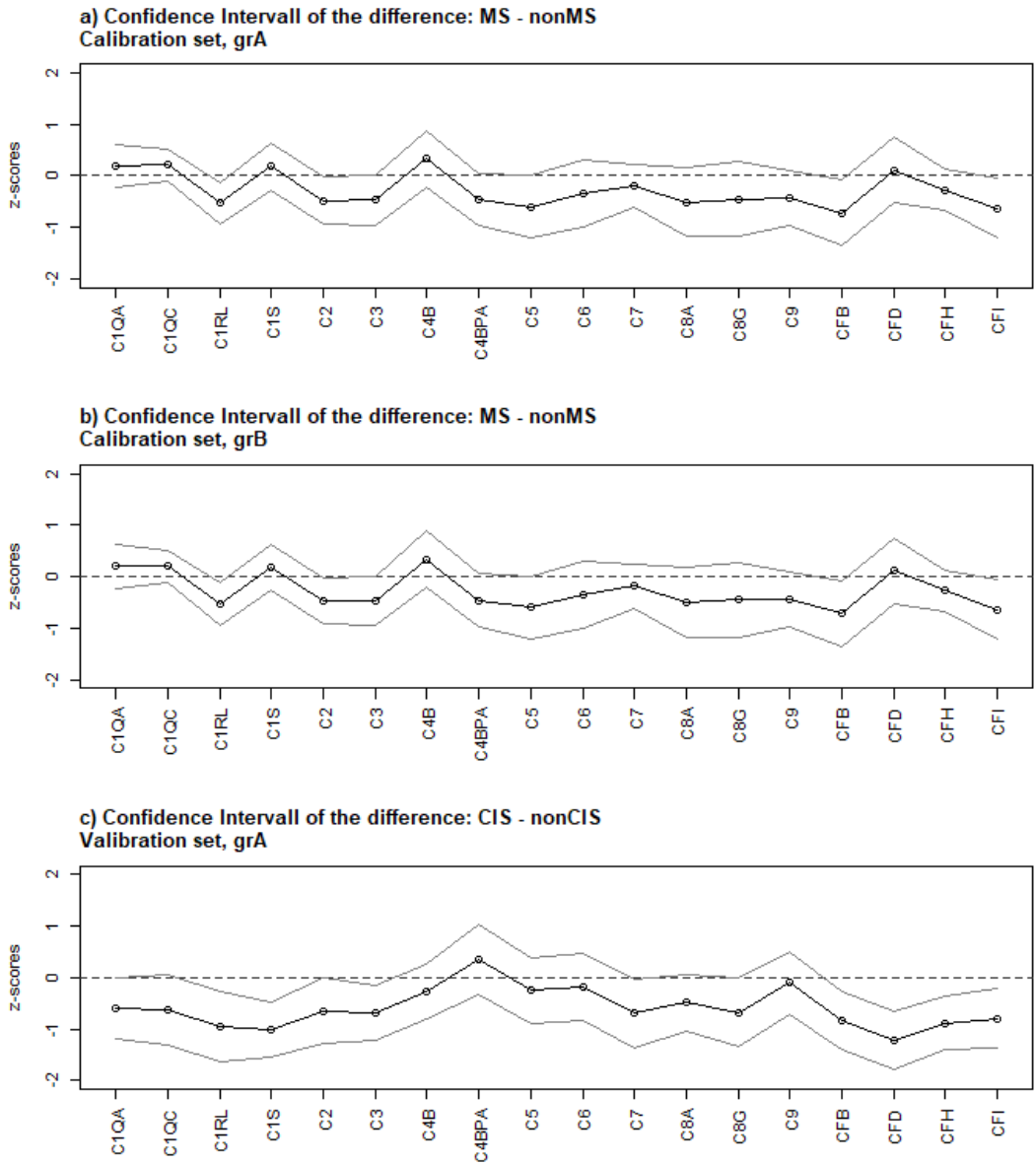

*Figure 2. **The MS dataset and the CIS datasets.** Confidence intervals of all complement proteins detected in the CSF, displaying the differences between MS/CIS and controls for all complement proteins. **(a)** The MS dataset group A, **(b)** The MS dataset group B. **(c)** The CIS dataset group A. The black line is the means, and the grey lines are the 95% confidence limits.*

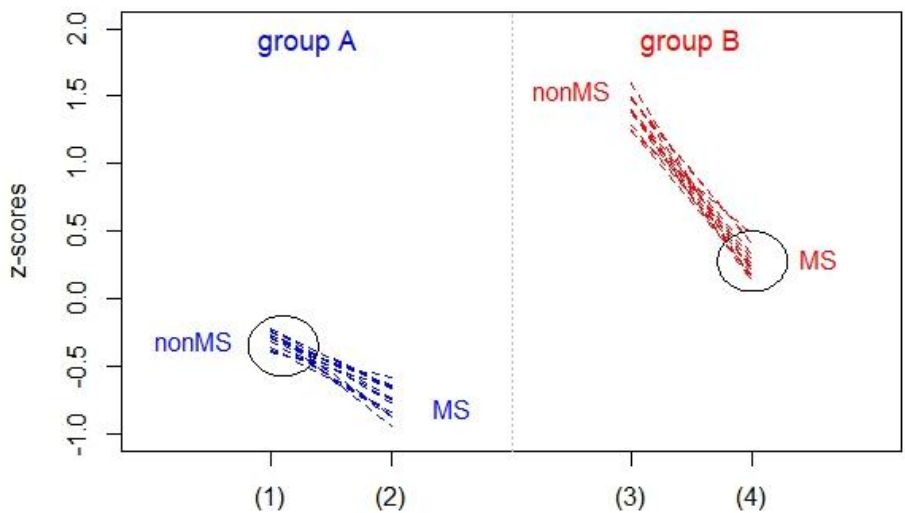

*Figure 3. The MS dataset.* The complement proteins C1RL, C2, C3, C4BPA, C5, C6, C7, C8A, C8B, C8G, C9, CFB, CFH, CFI presented as lines as means of groups and disease status. From the left: (1) group A nonMS, (2) group A MS, (3) group B nonMS, (4) group B MS. The figure reveals two notable patterns: group B has higher expression levels of these complement proteins than group A, and within each group, MS patients have lower expressions than non-MS patients. The most frequent combinations, both in this dataset and in general, are marked with circles: The circle to the left: "(1) group A nonMS", and the circle to the right "(4) group B MS".

We next applied GEM as it allows us to analyse all data across the groups while addressing both group identity and disease status.

**GEM**

GEM was applied to conduct patient stratification into groups A and B as one design variable, and nonMS-MS as another design variable. We also included gender and age at the lumbar puncture as design variables to omit any possible influence of these parameters when analysing nonMS compared with MS.

**GEM step 1, estimate effects and ER values of each design variable**

GEM step 1 of the MS dataset with four design variables is illustrated in **Figure 4**. GEM applied as a four-way model allows the effect MS to be considered without confounding the impact of group identity, gender, and age at lumbar puncture. Likewise for the analysis of the other design factors.

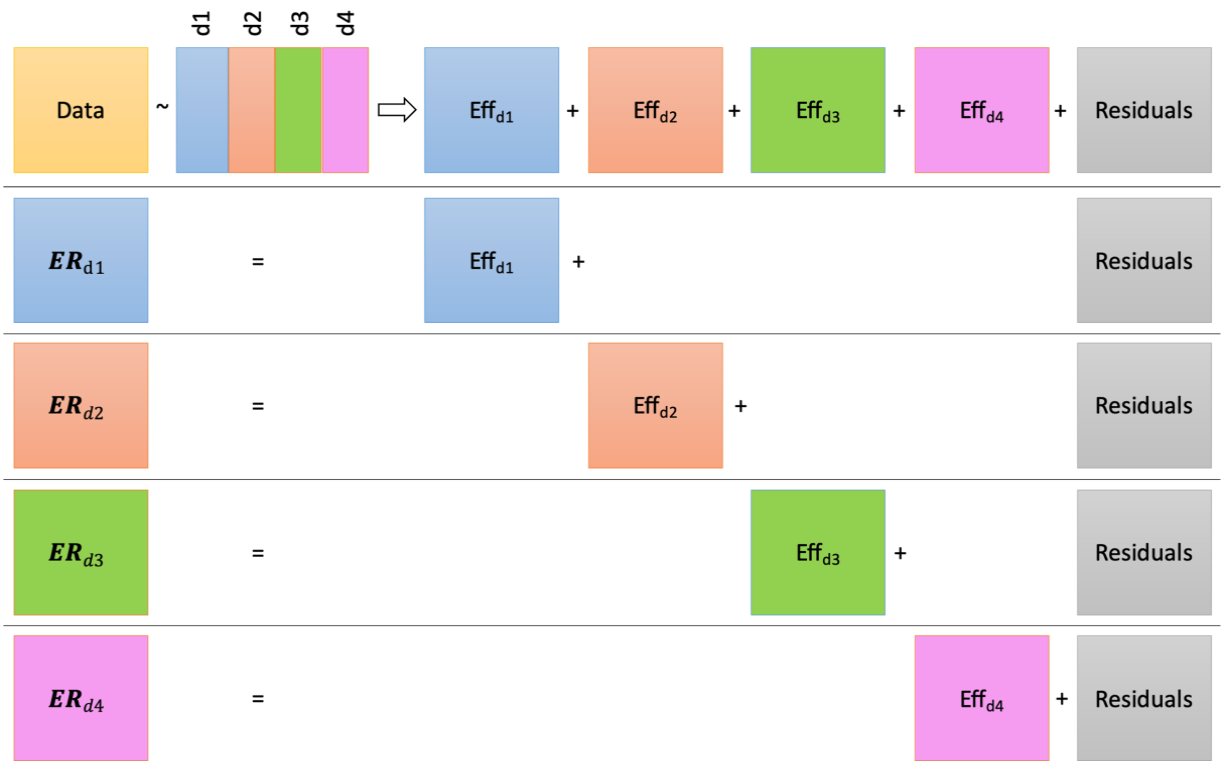

*Figure 4. **The MS dataset**. Illustration of GEM for the MS dataset. The upper road is the original model with all design variables included. **d1** is group identity (group A versus group B), **d2** is disease status (nonMS and MS), **d3** is gender (females versus males) and **d4** is the age at lumbar puncture. Lines 2, 3, 4, and 5 are ER values of each design variable d1, d2, d3 and d4, which are the effects of each design variable plus the residuals. The residuals are obtained from the whole model where all design variables are included (the upper line).*

To illustrate how GEM isolate the effects of disease status without confounding the impact of group, **Figure 5**, presents the data before and after GEM for the same complement proteins as displayed in **Figure 3**. The original data is displayed in **Figure 5a** (the same figure as **Figure 3a**). **Figure 5b** shows the ER values of the disease status (nonMS - MS). As shown in the figures there is a shift upwards for the complements in group A and a shift downwards for the complements in group B, when considering the ER values of the disease status (nonMS versus MS) in **Figure 5b** compared with the original data in **Figure 5a.** Thus, the effect group identity is omitted. The effects of MS compared with nonMS can thereby be analysed across all persons without confounding effects of group identity and disease status.

The same complement proteins are presented for CIS group A in **Figure 5c**. This reflected the same pattern of variation as observed within the groups of the MS datasets. These complement proteins were downregulated for CIS compared with nonCIS within group A. The observation that persons with CIS without CSF inflammation had downregulated complement proteins suggests that the complement proteins are downregulated at a very early stage of the disease.

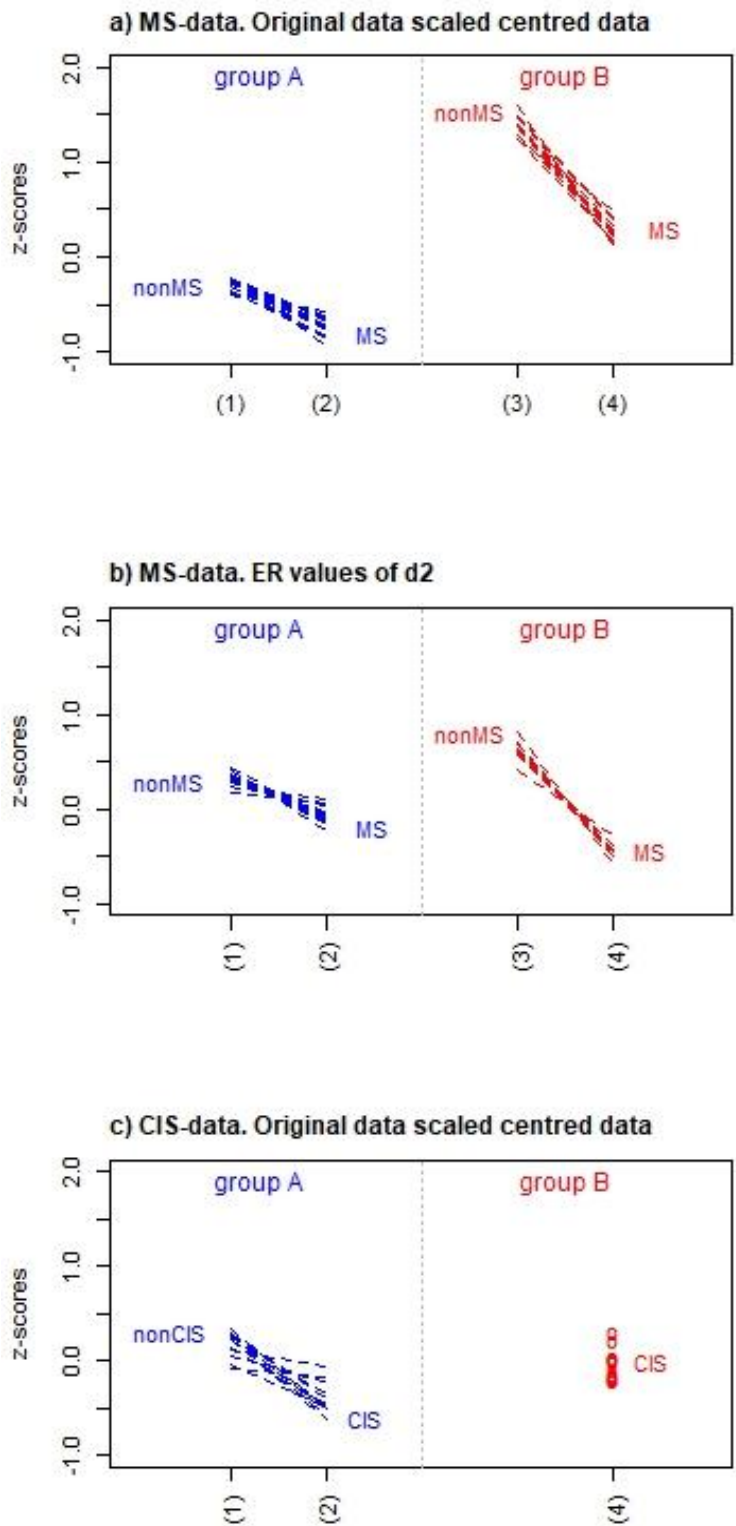

***Figure 5. The MS and the CIS datasets.*** *Illustration of GEM for all available complement proteins. Each line is means of one complement protein for the 4 categories, from the left: (1) group A nonMS, (2) group A MS, (3) group B nonMS, (4) group B MS. (**a**) The original data of the MS dataset, (**b**) ER values of the disease status (d2), and (**c**) the CIS dataset displaying the three available combinations: (1) group A nonCIS, (2) group A CIS, and (4) group B CIS.*

**GEM, step 2. Multivariate and univariate analysis of the ER values of each design variable**

Supervised multivariate analysis of all proteins by Partial Least Squares (PLS) [14, 15] applied in the GEM framework (GEM-PLS), for each of the four design variables in the MS dataset are displayed in **Figures 6**, **7** and **8**. The classification performances were high for the group identity and disease status, and somewhat lower for gender (**Figure 6a-c**), and the prediction result of the quantitative design variable age was high (**Figure 6d**) using two components in the PLS model.

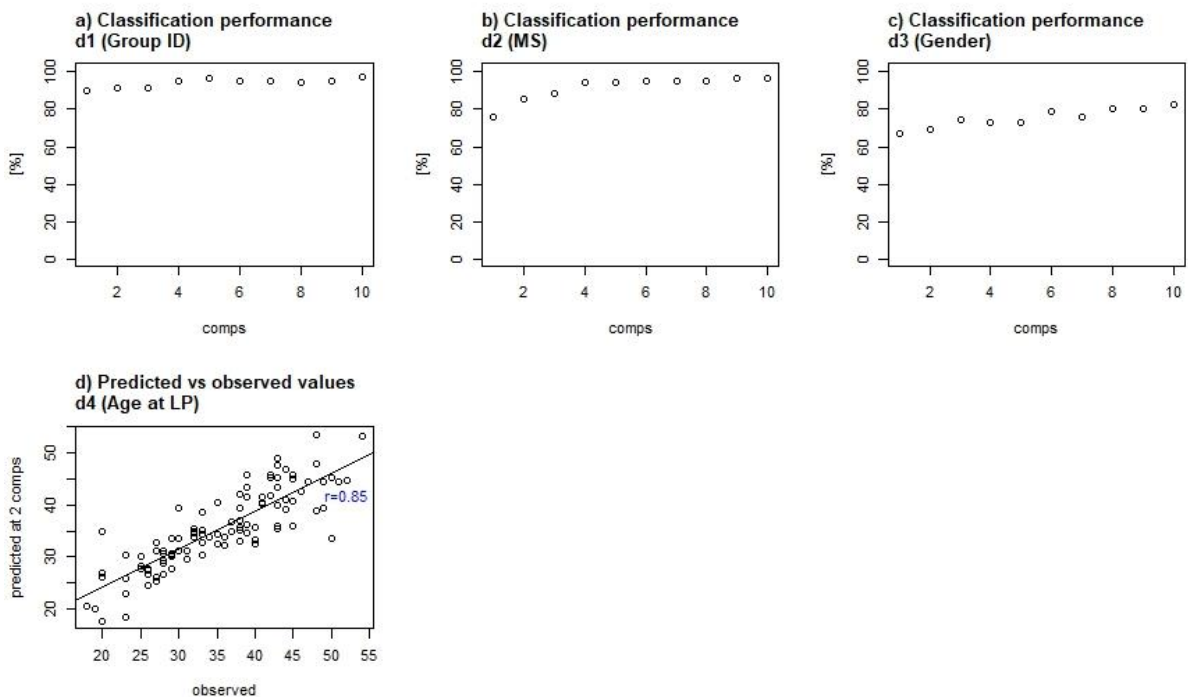

*Figure 6. The MS dataset (1). (a,b,c)* Classification performance of PLS discriminant analysis of *(a)* group identity (d1), *(b)* MS (d2), and *(c)* gender (d3). *(d)* Prediction performance of age at lumbar puncture (LP) (d4).

The GEM-PLS models of each design variable for the MS dataset are visualised for the two first PLS components by scores of the samples and corresponding loadings of the proteins. The two first components reflect well the design variables in each model: GEM-PLS on the design variable group identity (**Figure 7 a,b**), disease status (**Figure 7 c,d**), gender (**Figure 7 e,f**), and age at lumbar puncture (**Figure 7 g,h**).

**Figure 8** is a detailed view of the loading plot of the disease status where the complement proteins are highlighted. **Figure 8** shows that all the analysed complement proteins were strongly correlated in the loading plot, and located in the same direction as nonMS in the corresponding score plot (**Figure 7c**). Thus, the first component in PLS reflects a lower level of the complement proteins MS than for nonMS, which corresponds to the pattern of variation shown in **Figure 5**, when the effects of group identity is omitted.

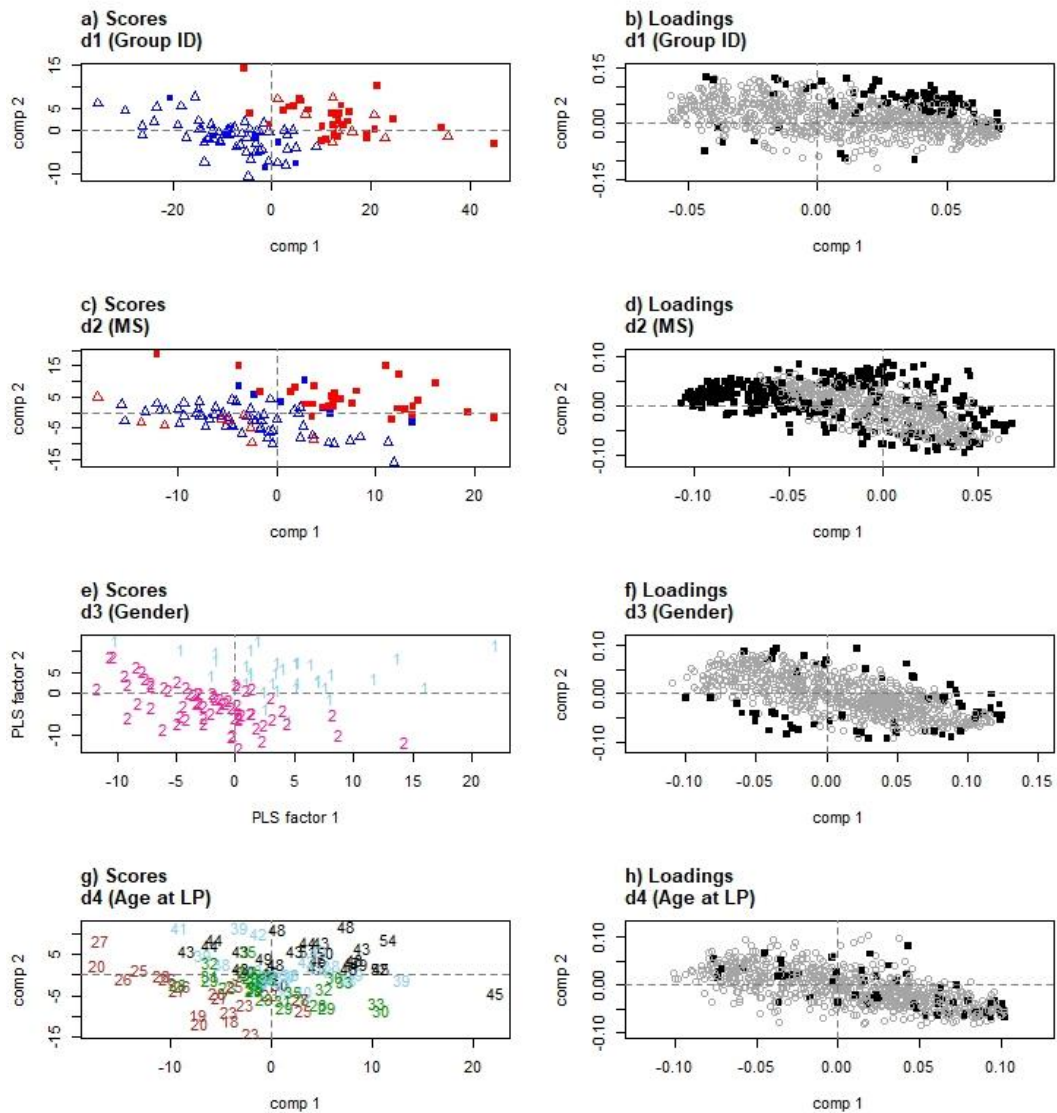

*Figure 7. **The MS dataset**. Scores and loadings of separate GEM-PLS models for each of the four design variables: (**a,b**) d1 (group identity) (**c,d**) d2 disease status (MS versus nonMS) (**e,f**) d3 (gender), (**g,h**) d4 (age at lumbar puncture). (**a,c,e,g**). Score plot of the samples, displaying (**a,b**) group A (blue) and group (B) for nonMS (open triangles) and MS (closed squares), (**e**) males (denoted 1 in light blue) and females (denoted 2 in pink), and (**g**) age at lumbar puncture (where the age is given in number and coloured by quantiles of age). (**b,d,f,h**) Corresponding loading plots of the proteins where all proteins are displayed in open grey circles and proteins significant by Jackknife [16] in the respective GEM-PLS with two components in the model are marked in black filled squares.*

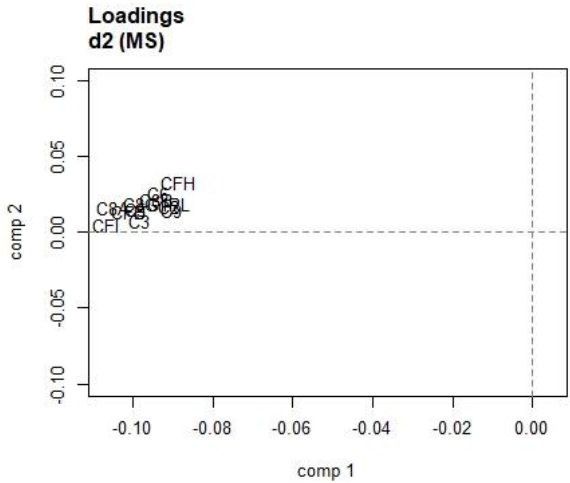

**Figure 8. The MS dataset.** Loadings of the GEM-PLS model applied on the design variable d2 (MS versus nonMS) displaying only the complement proteins.

Our interpretation is that complement deficiency may reflect a consequence of Epstein-Barr virus (EBV), as previously discussed [13]. EBV has for decades been suggested as a risk factor in MS – and recently shown to be a prerequisite for developing the disease [2, 3]. An experiment back in 1988 of serum incubated with purified EBV [17], suggested that EBV subvert complement- and cell-dependent host defense to provide survival value for itself. The complement proteins are also involved in non-immune processes during CNS development, progenitor proliferation, neural migration, and synaptic pruning from the embryonic stage to the adult stage [18]. Thus, the observed lower expression of complement proteins as considered within the group after stratification of the persons, may reflect a critical aspect in the molecular mechanism of MS. This observation may suggest that the normal development of the neuronal system may be restricted.

Another protein in focus for MS is transferrin (*TF*), the main transporter of iron, as brain iron homeostasis is crucial in the brain [19]. The switch between $Fe^{2+}$ ions as electron donors and $Fe^{3+}$ ions as electron acceptors is crucial for REDOX homeostasis and fundamental for the control of many biochemical reactions. The importance of iron in MS, especially linked to chronic inflammation, is increasingly studied in MS [20]. In addition to playing an important role in iron transport by reversibly binding $Fe^{3+}$, *TF* also has iron-independent functions. In a study of neuronal developmental parameters in a mouse neuroblastoma cell line, *TF* promoted neurite outgrowth and *TF* was suggested as a promising candidate to be used in regenerative strategies for neurodegenerative diseases [21].

In our study, we observed a dual pattern of variation of *TF* similar to that observed for most complement proteins. In the MS dataset (1), within groups A and B, *TF* was lower expressed in MS compared with nonMS, and *TF* was upregulated for group B compared with group A (**Figure 9 a**). Likewise, in the CIS dataset (2), *TF* was lower expressed for CIS compared with nonCIS within group A (**Figure 9 b**). Thus, among the groups without elevated CSF inflammation, CIS persons had lower expression of *TF* than nonCIS.

REDOX homeostasis is well known to play an important role in nearly all biological processes. Focus is often payed to the negative effects of excess oxidative stress, whereas later research has led to an improved understanding of the importance of Reactive Oxygen Species (ROS) signalling for many physiological processes including the development of the nervous system [22, 23]. ROS level is shown in isolated brain cells to play a key role in neuronal maturation [24].

The pattern of variation observed for MS and CIS represents an orchestrated pattern of variation that may shed light on underlying early events that may lead to MS. If the persons had not been stratified and considered as two groups as is here done, the pattern of variation that is here discovered would be unseen. Our interpretation, as presented in our former publication of this data [13], is that disturbed neural development may represent the earliest stage of MS development.

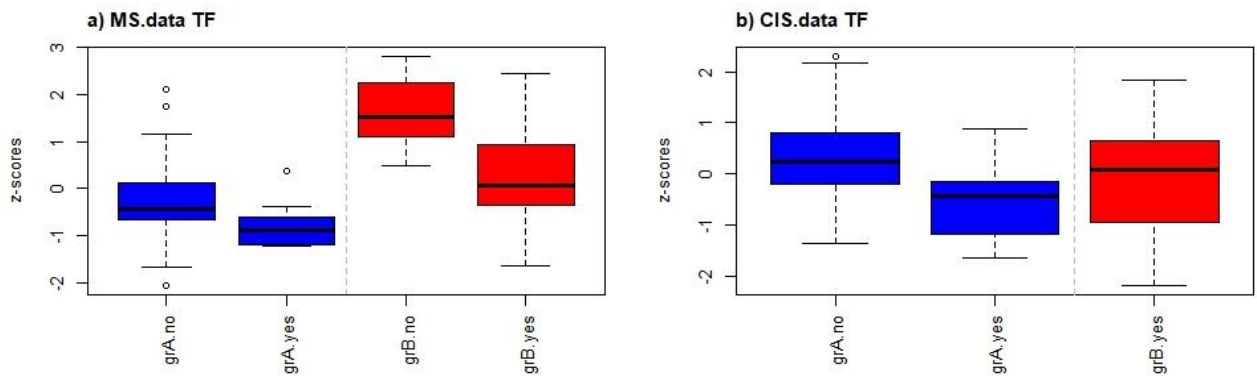

**Figure 9**. **The MS and the CIS datasets.** Boxplots of TF for (**a**) the MS dataset and (**b**) the CIS dataset. Within each dataset, the boxes represent group A (grA, in blue), and group B (grB in red) for persons without MS or CIS (no) and with MS or CIS (yes).

## Competing interests

We declare no conflicting interests